\documentclass{Rinton-P10x7}

\newcommand{\bx}{\raisebox{.75ex}{\framebox[2.5mm]{}}}
\begin{document}
\title{Black Hole Decay Rates in Large Extra Dimensions}
\author{Roberto Casadio}
\address{Dipartimento di Fisica, Universit\`a di Bologna,
and I.N.F.N., Sezione di Bologna, via Irnerio 46, I-40126 Bologna,
Italy\\
E-mail: casadio@bo.infn.it}
\author{Benjamin Harms}
\address{Department of Physics and Astronomy,
The University of Alabama, Box 870324, Tuscaloosa, AL 35487-0324,
USA\\E-mail: bharms@bama.ua.edu}
%
%%%%%%%%%%%%%%%%%%%%%%%%%%%%%%%%%%%%%%%%%%%%%%%%%%%%%%%%%%%%%%
% You may repeat \author \address as often as necessary      %
%%%%%%%%%%%%%%%%%%%%%%%%%%%%%%%%%%%%%%%%%%%%%%%%%%%%%%%%%%%%%%
%
\maketitle
\abstracts{We study the evaporation of black holes in space-times
with extra dimensions of size $L$.
We show that the luminosity is greatly damped when the horizon
becomes smaller than $L$ and black holes born with an initial size
smaller than $L$ are almost stable.
This effect is due to the dependence of both the occupation number
density of Hawking quanta and the grey-body factor of a black hole
on the dimensionality of space.}
\section{Introduction}
\setcounter{equation}{0}
If large extra spatial dimensions exist in nature, deviations from
Newton's law will be detected at the scale of the extra
dimensions.
Assuming that all of the matter described by the standard model
lives on a four-dimensional D3-brane, the form of Newton's law can
be obtained for a point-like mass by means of Gauss' law~\cite{arkani}.
For sufficiently large $L$ and $d$, the bulk mass scale $M_{(4+d)}$
(eventually identified with the fundamental string scale) can be as
small as $1\,$TeV. Since
\be
L\sim \left[{1\,{\rm TeV}/M_{(4+d)}}\right]^{1+{2\over d}}\,
10^{{31\over d}-16}\,{\rm mm}
\ ,
\label{tev}
\ee
requiring that Newton's law hold for distances larger than $1\,$mm
restricts $d\ge 2$ \cite{arkani,long}.
\par
As a black hole evaporates the topology of the horizon changes
from $S^2 \times R^d$ to $S^{2+d}$ when $R_H$ decreases from
$R_H>L$ to $R_H<L$.
For small black holes a complete description would provide an
explicit matching between the cylindrical metric (for $r\gg L$)
and the spherical metric (for $r\ll L$).
We approximate the metric in $4+d$ dimensions as
\be
\begin{array}{lr}
g_{tt}\simeq -1-2\,V(r)\ ,\ \ \ \ &
\ \ \
g_{rr}\simeq -g_{tt}^{-1}
\ ,
\end{array}
\label{met}
\ee
where $V(r)$ is an effective potential from which one
recovers the above-mentioned behaviors for $R_H \gg L$ and $R_H\ll L$
\cite{ch}.
The radius of the horizon, $R_H$, is then determined by
\be
V(R_H)=-{1/{2}}
\ .
\ee
\section{Black Hole Evaporation}
\setcounter{equation}{0}
%
%\label{evaporation}
%
%
{\em i) The Canonical Picture.}
When the size of the black hole is large compared to the extra
dimensions ($R_H \gg L$), the metric (\ref{met}) is approximately
cylindrically symmetric (along the extra dimensions).
The condition $R_H\gg L$ translates into
\be
M\gg m_p\,{L/\ell_p}\equiv
M_c \sim \left({L/1\,{\rm mm}}\right)\,10^{27}\,{\rm g}
\ .
\label{Mc}
\ee
\par
The energy loss per unit time for an evaporating black hole is
given by
\be
{dM\over dt}=-{F}_{(4+d)}^>\simeq-{F}_{(4)}^>
\ ,
\ee
where ${F}_{(D)}$ is the total luminosity as measured in $D$
space-time dimensions which can be approximated by employing
the canonical expression~\cite{hawking}
\be
\begin{array}{rcl}
{F}_{(D)}&\simeq&{\cal A}_{(D)}\, \sum_s\,
\strut\displaystyle\int_0^\infty
{\Gamma_s(\omega)\,\omega^{D-1} \over e^{\beta_H\,\omega}\mp
1}\,d\omega
\ ,
\label{L}
\end{array}
\ee
where $\Gamma$ is the grey-body factor and
$\beta_H=1/T_H^>\equiv 8\,\pi\,\ell_p\,M/m_p$ \cite{hawking}.
\par
\par
Although ordinary matter is confined on the D3-brane, a black hole
can emit particles via Hawking's process into all of the $3+d$
spatial directions of the bulk.
For $R_H \ll L$ the luminosity is given by Eq.~(\ref{L}) with
$D=4+d$ and $\beta_H=1/T_H^<\sim L\,(M/M_c)^{1/(1+d)}$~\cite{ch},
\be
{dM\over dt}=-{F}_{(4+d)}^< \sim {N_{(4+d)}\over
L^2}\,\left({M_c\over M}\right)^{2\over 1+d}
\ .
\label{L<}
\ee
The luminosity (\ref{L}) can also be written as
$F_{(D)}={N_{(D)}/ R_H^2}$ and a comparison with the analogous
quantity in four dimensions shows that a black hole of given ADM
mass $M<M_c$ emits much less in $4+d$ dimensions than it would do
with no extra dimensions.
\par\noindent
{\em ii) The Microcanonical Picture.}
The correct statistical mechanical description of
black holes utilizes the microcanonical ensemble \cite{r1,mfd}.
\par
For $R_H \gg L$ the topology of the horizon is ``cylindrical'',
and the Euclidean action is
$S^>_E \simeq 4\,\pi\,\left({M/{m_p}}\right)^2$.
The number density in the microcanonical enesemble for this case
is
\be
n^>(\omega)=\sum_{l=1}^{[[M/\omega]]}\,
{\exp\left[4\,\pi\,(M-l\,\omega)^2/m_p^2\right]
\over{\exp(4\,\pi\,M^2/m_p^2)}}
\ ,
\ee
where $[[X]]$ denotes the integer part of $X$.
In the limit $M \to \infty$, $n^>$ is equal to the canonical
ensemble number density used in Eq.~(\ref{L}).
After making the substitution $x = M-l\omega$, the decay rate in
four dimensions can be approximated by
\be
{dM\over dt}
={\pi^4\over{90}}\,{\cal A}_{(4)}\,\int_0^M
{(M-x)^3\,\exp\left(4\,\pi\,x^2/m_p^2\right)\,dx
\over{\exp\left(4\,\pi\,M^2/m_p^2\right)}}
\ ,
\ee
\par
\begin{figure}[t]
\raisebox{7pc}{${dM\over dt}$}
\epsfxsize=14pc
\epsfbox{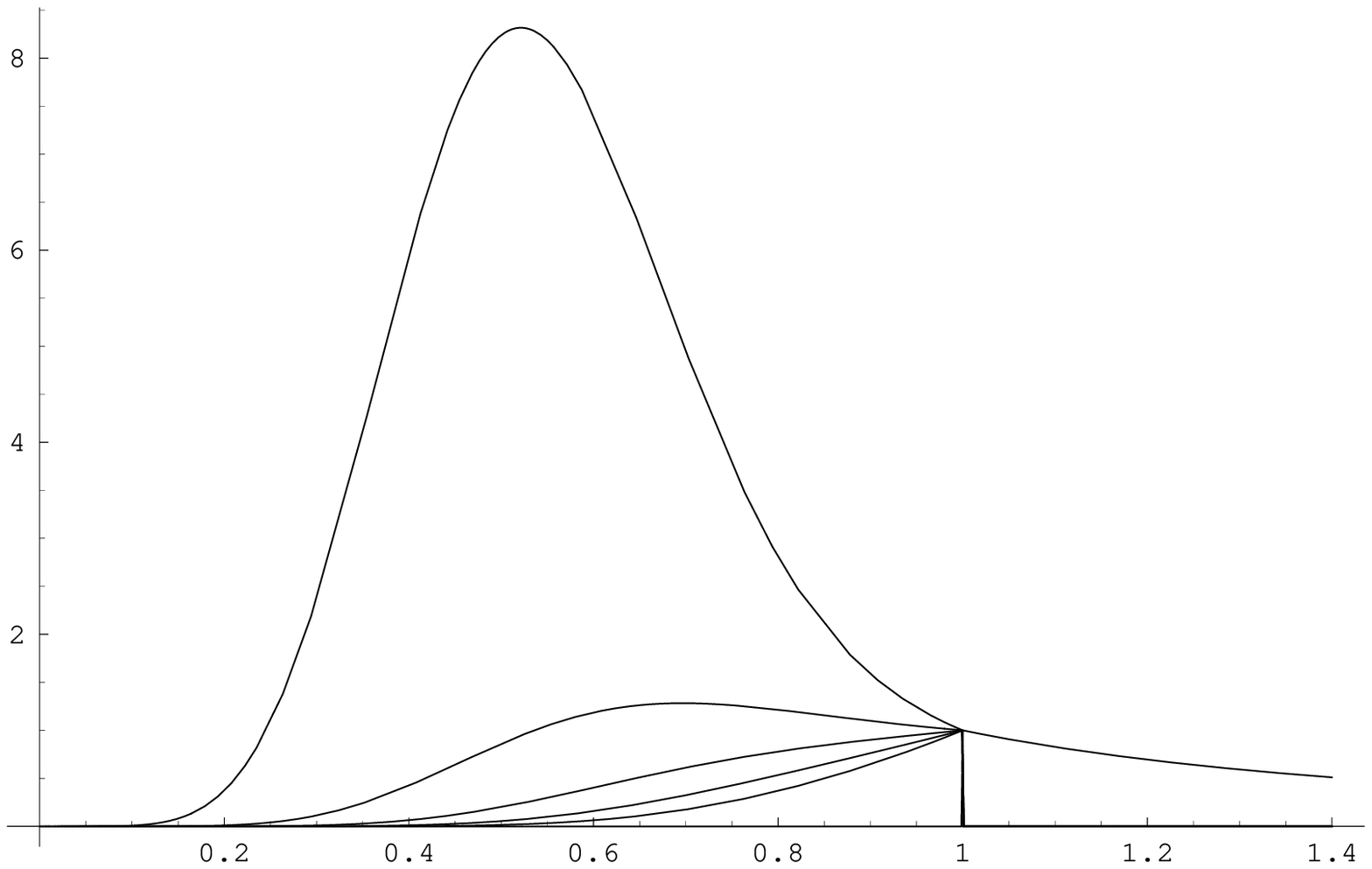}
\
\raisebox{7pc}{${V}$}
\epsfxsize=14pc
\epsfbox{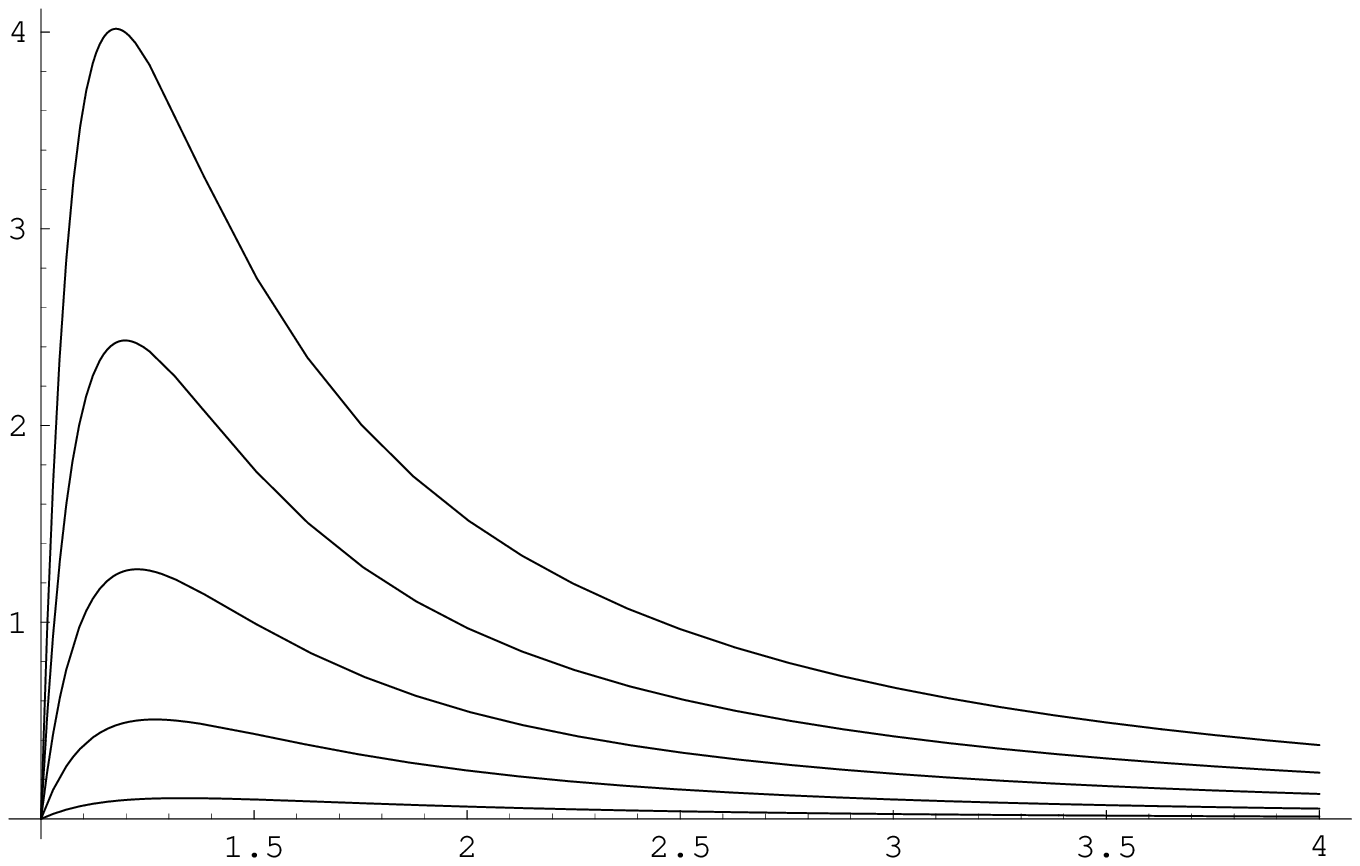}
\\
\hbox{\hspace{5truecm}{$M/M_c$}\hspace{6truecm}{$r/R_H$}}
\caption{Left) Decay rate for a black hole in increasing number of
extra dimensions ($d=0$ uppermost curve, $d=4$ lowest curve).
Vertical units are arbitrary.
Right) Potential $V$ in Eq.~(\ref{V}) for different numbers of extra
dimensions ($d=4$ uppermost curve, $d=0$ lowest curve).}
\label{microdMdt_s}
\end{figure}

For $R_H<L$ the Euclidean action is
\be
S_E^<=4\,\pi\,\left({L\over{\ell_p}}\right)^2\,
\left({M\over{M_c}}\right)^{(d+2)/(d+1)}
\ ,
\ee
where $M_c=m_p\,L/\ell_p$.
The number density for horizon radii of this size is
\be
n^<(\omega)=\sum_{l=1}^{[[M/\omega]]}\,
{\exp\left[4\,\pi\,
\left({M-l\,\omega\over{M_c}}\right)^{d+2\over
d+1} \,\left({L\over{\ell_p}}\right)^2\right]
\over{\exp\left[4\,\pi\,\left({M\over{M_c}}\right)^{d+2\over
d+1}\, \left({L\over{\ell_p}}\right)^2\right]}}
\ .
\ee
A plot of $dM/dt$ using this number density shows that the decay
rates have a maximum at $M \simeq M_c$ for all $d \ge 1$ and masses
less than $M_c$ have a significantly slower evaporation rate which
results in a longer lifetime (see left plot of
Fig.~\ref{microdMdt_s}).
\par\noindent
{\em iii) Angular Momentum Barrier to Decay.}
A scalar wave in $4+d$ dimensions satisfies the equation
\be
\bx\ \Phi=0
\ ,
\label{kg}
\ee
where \bx \ \ is the D'Alembertian.
For $L\ll R_H<r$ we can neglect the extra dimensions and simply
take the standard Schwarzschild line element on the brane,
\be
ds^2\simeq-\Delta_{(4)}\,dt^2+{dr^2\over \Delta_{(4)}}+
r^2\left(d\theta^2+\sin^2\theta\,d\phi^2\right)
\ ,
\ee
where $\Delta_{(D)}=1-\left({R_H/r}\right)^{D-3}$
\par
For $R_H<r\ll L$ one can analogously consider a spherically
symmetric black hole in $4+d$ dimensions \cite{myers}.
However, in order to take into account the fact that $d$ spatial
dimensions have size $L$, we shall instead use the following form
\be
ds^2\simeq-\Delta_{(4+d)}\,dt^2+{dr^2\over \Delta_{(4+d)}}+
r^2\,\left(d\theta^2+\sin^2\theta\,d\phi^2\right)
+r^2\,\sum_{i=1}^d\,d\phi_i^2
\ ,
\label{g4d}
\ee
where $R_H<r<L$ is the $4+d$ dimensional areal coordinate and
$dy^i\simeq r\,d\phi_i$ are cartesian coordinates in the extra
dimensions.
\par
We assume the scalar field $\Phi$ can be factorized according to
\be
\Phi=
e^{i\,\omega\,t}\,
R(r)\,S(\theta)\,e^{i\,m\,\phi}\,e^{i\,\sum\,n_i\,\phi_i}
\ ,
\ee
with $n_i$ positive integers, so that $\Phi$ satisfies periodic
boundary conditions at the edges of the bulk ($y_i=\pm L/2$).
The radial equation can be further simplified by defining a
tortoise coordinate $dr_*\equiv{dr/\Delta} $ and introducing a
rescaled radial function, $W\equiv r^{1+d/2}\,R$, we have
\be
\left[-{d^2\over dr_*^2}+V\right]\,W=\omega^2\,W
\ ,
\ee
where the potential is given by
\be
\begin{array}{rcl}
V=\left[(1+d)\,\left(1+{d\over
2}\right)+ A\right]\, {\Delta\over{r^{2}}}
-\,\left(1+{d\over 2}\right)^2\,{\Delta^2\over r^2}
\ .
\label{V}
\end{array}
\nonumber
\ee
A plot  of the potential shows an angular momentum barrier
which increases with increasing $d$ (see right plot of
Fig.~\ref{microdMdt_s}, in which units are so chosen that
$\omega=T_H=1$).
From this plot one can estimate the (frequency dependent)
suppression factors with respect to the purely four-dimensional
case (for which $\Gamma\sim 1$) by making use of the W.K.B.
approximation for the transmission probability
\be
\Gamma(\omega)\sim\exp\left(-2\,
\int dr\,\sqrt{\left|V-\omega\right|}\right)
\ ,
\ee
where the integral is performed between the two values
of $r$ at which $V=\omega$ (for $\omega$ smaller than the maximum
of $V$), obtaining $\Gamma\simeq 1$, $0.73$, $0.22$, $0.05$ for
$d=1,\ldots,4$ and $\omega=T_H$~\cite{ch}.
\section{Conclusions}
\par
The analysis of the black hole decay rate in both the canonical
and microcanonical pictures shows that the presence of large extra
dimensions would slow the decay rate, and that the decay rate
decreases with increasing number of extra dimensions. The source
of this decrease in decay rate is the dependence of the number
density and the horizon area on the number of extra dimensions. In
addition to these effects the size of the angular momentum barrier
increases with the number of extra dimensions, further decreasing
the decay rate. This result suggests that, even without taking
backreaction effects into account, black holes with horizon radii
less than the size of the extra dimensions are quasi-stable.
\section{Acknowledgments}
This work was supported in part by the U.S. Department of Energy
under Grant no.~DE-FG02-96ER40967 and by NATO grant
no.~CRG.973052.

\end{document}